\documentclass[%
 reprint,
 amsmath,amssymb,
 aps,
floatfix,
]{revtex4-2}

\usepackage{hyperref}
\hypersetup{
    pdfproducer={},  
}

\usepackage{graphicx}
\usepackage{dcolumn}
\usepackage{bm}

\begin{document}


\title{On age-specific selection and extensive lifespan beyond menopause}

\author{Tin Yau, Pang}
 \email{pang@hhu.de}
\affiliation{%
 Computational Cell Biology Group \\
 Heinrich Heine University D\"usseldorf
}%

\date{\today}

\begin{abstract}
Extensive post reproductive lifespan (PRLS) is observed only in a few species, such as humans or resident killer whales, and its origin is under debate. To explain PRLS, hypotheses like mother-care and grandmother-care invoke strategies of investment---provision to one\textsc{\char13}s descendants to enhance its overall reproductive success. The contribution of an investment strategy varies with the age of the caregiver, as the number of care-receiving descendant changes with age. Here we simulate an agent based model, which is sensitive to age-specific selection, to examine how the investment strategies in different hypotheses affect survival and reproduction across different stages of life. We showed that extensive PRLS emerges if we combine multiple investment strategies, including grandmother-care but not mother-care, which allow an individual to have an increasing contribution as it ages. We also found that, if mother-care is further introduced to a PRLS-enabling strategy, it will let contribution at mid-life to substitute contribution at late-life, which consequently terminate extensive PRLS.
\end{abstract}

\maketitle


\begin{table}[b]
\caption{\label{tab:symbols}
List of strategies of investment in offspring and grand-offspring.
}
\begin{ruledtabular}
\begin{tabular}{|p{1.1cm}|p{2.8cm}|p{4.2cm}|}
\textrm{Symbol}&
\textrm{Name}&
\textrm{Description}\\
\colrule

\hline
NULL  &  basic null model  &  original model, without extra interaction \\
\hline
M  & mother-care & the ad-hoc death rate of an agent at life stage $i \in {0,1}$ is reduced by 10 folds if its mother is present \\
\hline
GM  & grandmother-care & the ad-hoc death rate of an agent at life stage $i \in {0,1}$ is reduced by 10 folds if its maternal grandmother is present \\
\hline
LTr  & reproduction-enhancing long-term-care & the ad-hoc birth rate of an agent at life stage $i \geq 8$ becomes $min(1, 2r_i)$ if its mother is present \\
\hline
LTs  & survival-enhancing long-term-care & the ad-hoc death rate of an agent at life stage $i \geq 2$ is reduced by 10 folds if its mother is present \\
\end{tabular}
\end{ruledtabular}
\end{table}

\begin{figure}[b]
\includegraphics[width=0.49\textwidth]{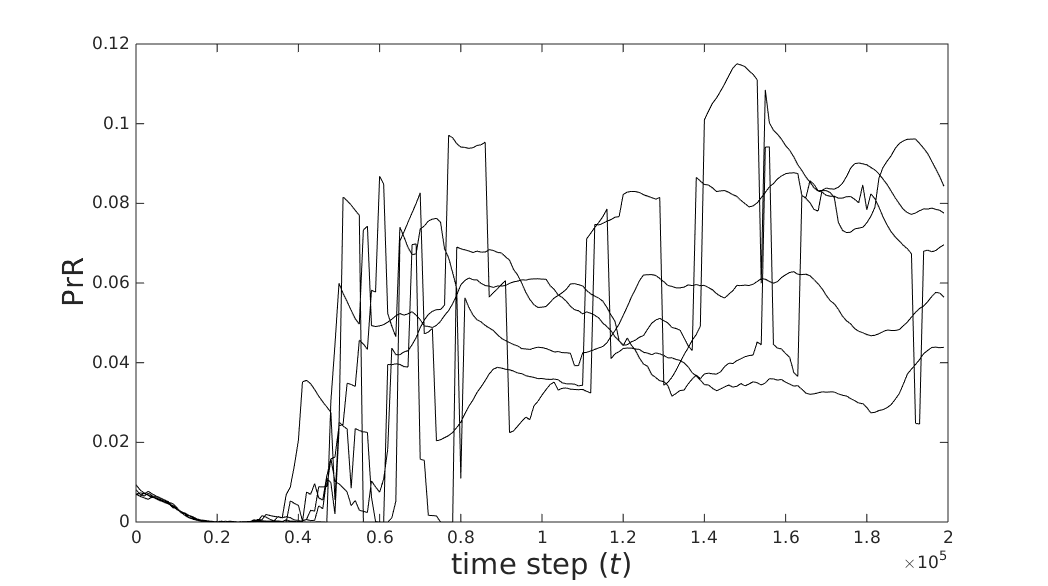}
  \caption{\label{fig:relaxationNULL} Post-reproductive representation (PrR) of the female individuals in the simulated populations of condition NULL. It shows that the relaxation process ends within 100,000 steps.
}
\end{figure}

\begin{figure}[b]
\includegraphics[width=0.49\textwidth]{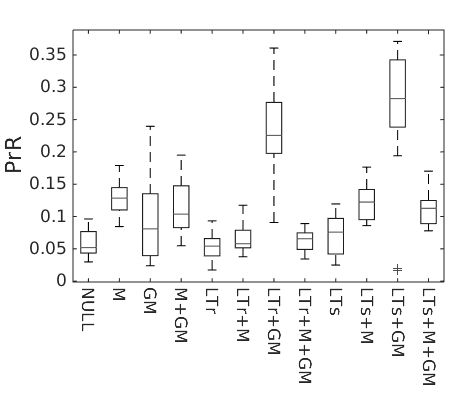}
  \caption{\label{fig:xxx12PrR} Distribution of PrR in different conditions.
}
\end{figure}

\begin{figure}[b]
\includegraphics[width=0.49\textwidth]{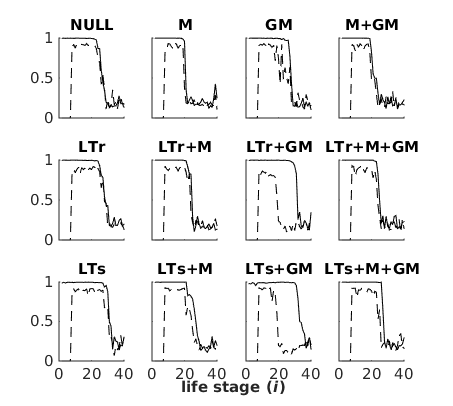}
  \caption{\label{fig:figSR} Female intrinsic rate of survival $s_i$ (solid curve) and reproduction $r_i$ (broken curve). Each curve is an average over every chromosome at the end of five different simulations.
}
\end{figure}

\begin{figure}[b]
\includegraphics[width=0.49\textwidth]{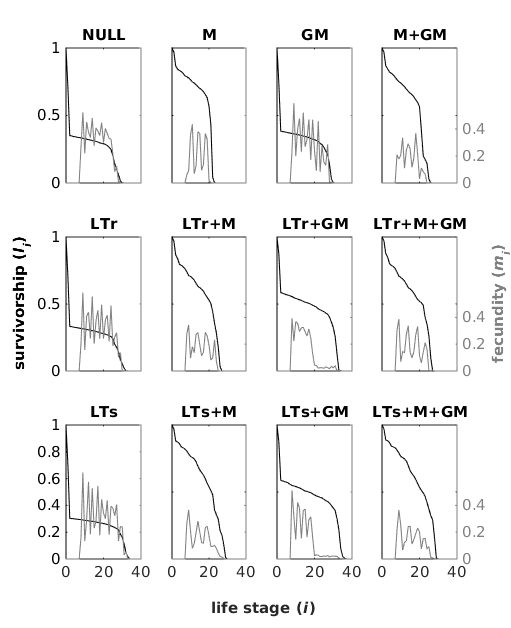}
  \caption{\label{fig:xxxLifeTable} Survivorship $l_i$ (black) and individual-fecundity $m_i$ (grey) of female individuals. Each curve is inferred from the statistics sampled at the end of five different simulations.
}
\end{figure}

\begin{figure}[b]
\includegraphics[width=0.49\textwidth]{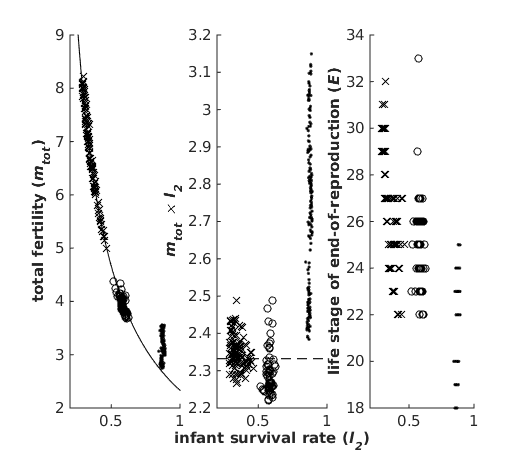}
  \caption{\label{fig:F_ML_EoR_l2} (a) Fertility, defined as total fecundity $m_{tot}$, (b) the product of fertility and infant survival rate, and (c) the end-of-reproduction life stage ($E$) plotted against the infant survival rate, defined as $l_2$---the probability to survival to stage 2. Circle markers are data from conditions that have extensive PRLS, dots are conditions that involve mother-care, and crosses are the other tested conditions. The solid curve in (a) is $y=2.33/x$. The broken curve in (b) is $y=2.33$.
}
\end{figure}

\begin{figure}[b]
\includegraphics[width=0.49\textwidth]{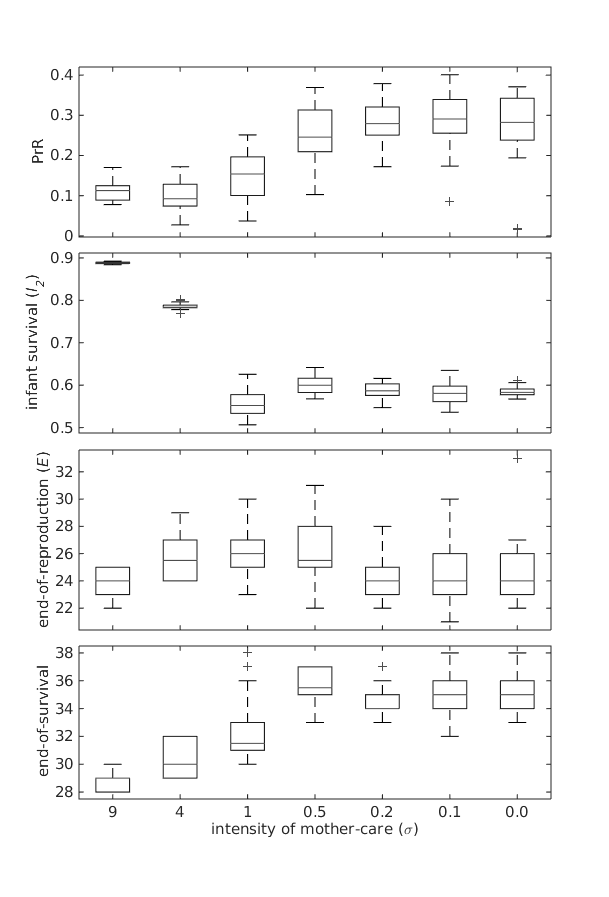}
\caption{\label{fig:xxxtransition} Distribution of PrR, survivorship at 2nd life stage ($l_2$), end-of-reproduction life stage ($E$), and end-of-survival life stage of the intermediate conditions between LTs+M+GM ($\sigma=9$) and LTs+GM ($\sigma=0$). In these conditions, grandmother-care (GM) and survival-enhancing long-term care (LTs) are present. We tuned the intensity of mother-care by varying $\sigma$, as the ad-hoc death rate of a dependent infant is reduced by a factor $1/(1+\sigma)$ if its mother is present.}
\end{figure}

\begin{figure}[b]
\includegraphics[width=0.49\textwidth]{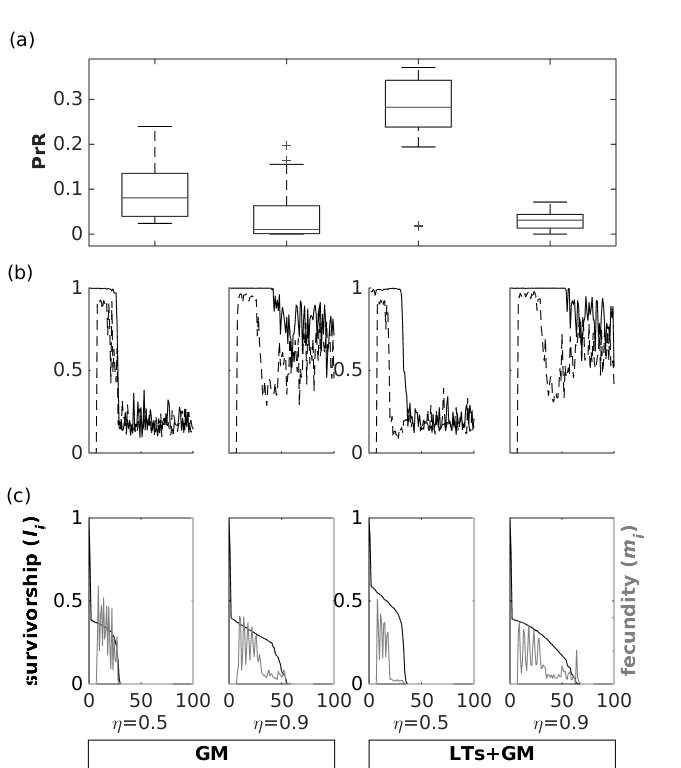}
  \caption{\label{fig:boxplot_sensitivity} (a) Distribution of PrR, (b) intrinsic rate of survival $s_i$ (solid curve) and reproduction $r_i$ (broken curve) averaged over every chromosome at the end of simulations, and (c) survivorship $l_i$ (black) and individual-fecundity $m_i$ (grey) inferred from the statistics sampled at the end of the simulations, for the conditions GM and LTs+GM, with beneficialness-to-deleteriousness ratio for mutation, $\eta$, equals 0.5 (default) and 0.9.
}
\end{figure}

Evolution selects for individuals based on their reproductive success, which depends not only on reproductive rate, but also on time and effort invested in the future generations \cite{Lee2003Rethinking}. The theory of ageing predicts differential selection on the rate of survival and reproduction at different stages of life: because there are fewer old individuals than younger ones, the strength of selection on age-specific loci gets weaker with increasing age. Therefore, deleterious mutations affecting early survival tend to be removed by purifying selection, whereas those affecting late survival tend to accumulate \cite{Charlesworth2000,haldane1942new,Hamilton1966,medawar1952unsolved,Rose1990}. Likewise, efficacy of investment is also age-specific, as the expected number of offspring and grand-offspring changes with age.

In most animal species, the reproductive lifespan coincides with the somatic lifespan. There are a few exceptions, such humans and resident killer whales, whose female individuals experience extensive post reproductive lifespan (PRLS)  \cite{williams1957pleiotropy,Hamilton1966,medawar1952unsolved,olesiuk1990life}. Many researchers believe that investment in future generations is the primary cause of extensive PRLS. Human babies are born with head size close to the limit of safe delivery \cite{abitbol1996birth}, yet their brain needs further development before becoming capable of independent survival. Accordingly, the death of a mother reduces the survivorship of its newborns \cite{sear2008keeps}, and because of the higher risk of late-life pregnancy, terminating reproduction and investing in the dependent offspring or grand-offspring may be a better strategy \cite{williams1957pleiotropy,Penn2007}. Alternatively, the investment on adult offspring may also be a contributing factor---adult male resident killer whales have higher survival \cite{Foster2012Adaptive}, or adult female humans have higher fertility \cite{sear2008keeps}, if their mother is present. Therefore, we also examined the hypotheses of investment in adult offspring that raise their survival or reproduction rate.

Recent theoretical and in-silico studies investigated the link between mother-care / grandmother-care and PRLS \cite{Kim2014Grandmothering,Kim2019,Kim2012Increased,Shanley2007Testing,Kachel2011Grandmothering} and found mixed support (see Ref \cite{Croft2015Evolution} for review). Some of these models may have been oversimplified, such as assuming a non-evolvable reproduction lifespan \cite{Kim2014Grandmothering}. Others may impose a-priori trade-off functions, e.g., trade-off between fertility and survival, to prevent undesirable behaviours of the model, such as ``cost free evolution'' that drives the agents ``toward greater and greater longevity'' \cite{Kim2019}.

Here we simulate a model of adaptive agents modified from Ref \cite{Sajina2016} to test these hypotheses. This model does not employ any a-priori trade-off functions to constrain the survival or reproduction rates, instead each agent is parametrized by evolvable, age-specific survival and reproduction rates encoded in its genome. The deleterious-prone mutation drives the value of every parameter towards zero and prevents an agent from evolving towards unbounded reproductive or somatic longevity. Parameters critical to fitness are, however, maintained at high value by purifying selection.

Our model considers a population that evolves through steps of time. An agent can be male or female. The integer index $i$, $i \in {0,...,101}$, labels the life stage of an agent, and increases by 1 in each time step. The agent dies if it reaches $i=101$. $s_i$ and $r_i$ of an agent respectively denote its intrinsic survival and reproduction rate of stage $i$. An individual at $i \in {0,1}$ is unweaned and depends on external provision, we set $s_i$ in these stages to be non-evolvable and equal 0.9 to represent its weakness. Rapid extinction may occur if we set a lower survival rate, e.g., 0.7. Reproduction starts at $i=8$. These intrinsic survival and reproductive rates are encoded in the genome and evolvable, they form 192 evolvable parameters that govern the behaviour of an agent.

The genome of an agent is diploid and has two chromosomes. A chromosome is represented by a sequence of loci that correspond to the parameters. There is a set of parameters for male, another set for female, and so a chromosome has $384=192 \times 2$ loci. The parameters for male have no effect if the agent is female, and vice versa. The intrinsic $s_i$ (and $r_i$) of an agent is the average value of the corresponding locus on the two chromosomes.

A simulation starts with 1,000 agents that have uniform genome. Each agent has its gender randomly assigned, and its life stage randomly assigned from the range $0 \leq i \leq 50$. Initial parameters include $s_i=0.97$ for all $i$s, $r_i=0.4, 8 \leq i \leq 16$, and $r_i=0.1, i \geq 17$. These initial values appear arbitrary, they are parametrized in this way because smaller values sometimes lead to rapid extinction. We emphasize that selection will ultimately determine their value after the relaxation process in the simulation. In each step, 1000 unit of resource is replenished, each agent consumes one unit, and the surplus resource will not be transferred to the next step. Let $N_t$ be the number of agents at step $t$. When $N_t \leq 1000$, the probability for an agent at stage $i$ to survive to the next step is $s_i$. A female at life stage $i \geq 8$ gives birth to one child with probability $r_i$. To give birth, it pairs up with a random mature male from the population, weighted by their reproduction rates. The gender of the newborn agent is randomly assigned. It receives one arbitrary chromosome from each parent, which thereafter undergoes crossover and mutation. When $N_t > 1000$, famine occurs and the chance for survival and reproduction is compromised. For simplicity, the ad-hoc reproduction rate of all agents is set to be 0. The intrinsic death rate of an agent, $1-s_i$, is magnified by an exponential factor $3^{T_f}$, where $T_f$ is the number of consecutive stages that famine has lasted \cite{Sajina2016}. The ad-hoc survival rate, thus, is $1-3^{T_f}(1-s_i)$.

The possible values at a locus is discrete, which are 0, 0.20, ..., 0.60, 0.80, 0.82, ..., 0.88, 0.90, 0.91, ..., 0.98, 0.99, 0.999 for $s_i$, and 0.0, 0.1, ... , 0.9, 1.0 for $r_i$. The values of $s_i$ have uneven intervals, because this allows a locus to drop from large value to 0 quickly when it is not selected for and saves computational resource. During crossover, the two chromosomes swap their segments. There is 10\% chance for a position between two loci to be a crossover breakpoint, which serves as start and end of crossover segments. After crossover, each locus is mutated at a rate 0.025. We define the beneficialness-to-deleteriousness ratio of mutation, $\eta$, to be 0.5. A locus chosen to mutate has 50\% chance to decrease by one level, $50\% \times \eta = 25\%$ chance to increase by one level, and $50\% \times (1-\eta) = 25\%$ chance to have no change.  This deleterious-prone nature of mutation is consistent with experimental observations \cite{Sarkisyan2016}, it makes a locus not selected for to have close-to-zero value.

We considered several strategies of investment in descendants by females. The condition ``NULL'' denotes the model without any add-on interactions. ``Mother-care'' (M) (``grandmother-care'' (GM)) allows the reduction of death rate of an unweaned dependent infant by 10 fold if its mother (maternal grandmother) is alive. We also imposed two types of ``long-term-care'' (LT): an independent agent has (a) a higher reproduction rate (LTr), or (b) a higher survival rate (LTs), if its mother is alive. See Table \ref{tab:symbols} for details of these investment strategies. These conditions can be combined, e.g., with condition M+GM, the ad-hoc death rate of a dependent infant is reduced by 10 fold if its mother is alive, and by another 10 folds if its maternal-grandmother is alive.

We calculated the survivorship, $l_i$, and individual-fecundity, $m_i$, of the female individuals of a population to infer their reproduction and somatic longevity. The survivorship and fecundity at time $t$ are calculated from the statistics of the population sampled within the period $[t-10000, t]$. Specifically, $l_i$ is the probability for a newborn to survive to stage $i$, and $m_i$ is the average reproductive output---the chance to give birth---of an individual at stage $i$ \cite{Begon1996}. We quantified PRLS using post-reproduction time (PrT) and post-reproduction representation (PrR) \cite{Levitis2011Measure}. The average remaining lifespan at stage $i$, $e_i$, is defined as \[e_i = \frac{\sum_{k=i}^{\infty} (k-i)(l_k - l_{k+1})}{\sum_{k=i}^{\infty} l_k - l_{k+1}} \]
Let $B$ and $E$ be the smallest integers that respectively satisfy $\sum_{i=0}^B m_i \geq 0.05 \sum_{i=0}^\infty m_i$ and $\sum_{i=0}^E m_i \geq 0.95 \sum_{i=0}^\infty m_j$. $B$ and $E$ represent the stage of begin-of-reproduction and end-of-reproduction, respectively. $E$ is also called reproductive longevity. Let us define $PrT = e_E$, the expected lifespan after the end-of-reproduction, which is intuitive but vulnerable to statistical noise. \textit{Croft et al.} pointed out that, a tiny number of exceptionally long-living individuals in a sample could lead to a high PrT, and hence a false-positive indication of extensive PRLS \cite{Croft2015Evolution}. Let us also define \[ PrR = \frac{l_E e_E}{l_B e_B} \] which is less intuitive but statistically robust. PrR is $\ll 0.2$ for species without extensive PRLS, e.g., 0.02 for wild Chimpanzee, and higher otherwise, e.g., 0.22 for resident kill whales, 0.32-0.71 for different human samples \cite{Croft2015Evolution}. We used 0.20 as the cutoff PrR for extensive PRLS.

Simulation of the NULL condition showed that the relaxation time of the model is well below 100,000 time steps (Fig. \ref{fig:relaxationNULL}). We also simulated different combinations of investment strategies, which include the conditions M, GM, M+GM, LTr, LTr+M, LTr+GM, LTr+M+GM, LTs, LTs+M, LTs+GM, LTs+M+GM (see Table \ref{tab:symbols} for definition of symbols). A condition is simulated 5 times, each lasted for 200,000 steps. Starting from the 100,000-th step of each simulation, we calculated the survivorship and fecundity every 20,000 steps to infer PrR and other properties of the population for further analysis. We only found two conditions with PrR unambiguously $\geq 0.20$ and have extensive PRLS emerges: LTr+GM and LTs+GM (Fig. \ref{fig:xxx12PrR}).

As predicted by the theory of ageing, strong selection on survival at early ages drives $s_i$ towards one (Fig. \ref{fig:figSR}). High infant death, however, leaves a sharp kink on the curve of survivorship at early stages in several conditions like NULL (Fig. \ref{fig:xxxLifeTable}), which is also observed in human and whale populations (see, e.g., Ref \cite{Croft2015Evolution}). In NULL, the survivorship drops by 60\% within the first two life stages and seems incongruent with the fixed 0.1 intrinsic death rate of dependent infant. This is because the population size occasionally increases beyond the available resource and thereby leads to famine and an elevated ad-hoc death rate.

Mother-care (M) effectively protects dependent infants, the sharp kink of survivorship near $i=2$ hence disappears, and more newborns can survive to adulthood. This consequently weakens the selection on fertility, and results in a shorter reproductive and somatic lifespan compared with NULL (Fig. \ref{fig:xxxLifeTable}). Driven by the benefit of caring for the last-born, the average lifespan after menopause increases: compared with the PrT of NULL ($1.19\pm0.39$), PrT of M is higher ($1.73\pm0.28$) and closer to 2---number of stages of unweaned dependent infanthood. As opposed to mother-care, grandmother-care alone (GM) can only very slightly mitigate the low survivorship of dependent infants (Fig. \ref{fig:xxxLifeTable}), because the chance for a dependent infant to have a living grandmother is much lower than mother.

Investment in the descendants affects the infant survival and adult fertility. Let us quantify infant survival by $l_2$, and fertility by total lifetime fecundity, $m_{tot} = \sum_i m_i$. Fertility and infant survival can be well summarized by the equation \[ m_{tot} = \frac{2.33}{l_2} \] (see cross and square-markers in Fig. \ref{fig:F_ML_EoR_l2}), except for conditions involving mother-care (see dot-markers in Fig. \ref{fig:F_ML_EoR_l2}). The value 2.33 is the average $m_{tot} \times l_2$ for conditions not including mother-care. It can be roughly interpreted as the average number of offspring per adult female individual, which is indeed the equilibrium value of a ``tug-of-war'' process. On the one hand, agents with a low fertility tend to be out-competed, which effectively results in an upward force on fertility. On the other hand, a large number of agents with a high fertility may lead to more frequent famine and death that nullifies this upward selection force, and the deleterious mutation acts like a downward force. The tug-of-war of these forces defines the equilibrium fertility. Nonetheless, conditions that involve mother-care are outliers to this equation. Mother-care dramatically enhances infant survival, making the resource of the population more stressful. This affects the pattern of famine and population dynamics, and hence the equilibrium of fertility. Moreover, while the mapping between fertility and infant survival is well-behaved and collapses onto a line, the mapping between fertility and reproductive lifespan is far from a line, but still strongly correlated (spearman correlation: $\rho=0.8474, p<2.2 \times 10^{-16}$).

Extensive PRLS emerges only in strategies LTr+GM and LTs+GM, with PrR equals $0.23 \pm 0.06$ and $0.28 \pm 0.09$, and PrT equals $5.77 \pm 1.38$ and $7.26 \pm 2.07$, respectively. What makes these two conditions stand out from the rest? The efficacy of mother-care, long-term-cares and grandmother-care scale with age in different ways. Let us quantify the efficacy of an investment strategy at stage $i$ by $c_i$, the average number of care-receivers. $c_i$ for mother-care, reproduction / survival-enhancing long-term-care, and grandmother-care are, respectively, 
\[ c_i^M = m_{i-2}l_1 + m_{i-1}l_0 \]
\[ c_i^{LTr} = \sum_{j=8}^{i-9} m_jl_{i-j-1} \]  
\[ c_i^{LTs} = \sum_{j=8}^{i-2} m_jl_{i-j-1} \]
\[ c_i^{GM} = \sum_{j=8}^{i-1} \sum_{k=0}^{i-1} \frac{m_j}{2} l_k m_k (l_1 \delta_{i-j-k-3} + l_0 \delta_{i-j-k-2}) \]
Here, $\delta_{x}$ is the Kronecker delta, and the factor 2 in the denominator of $c_i^{GM}$ accounts for grand-offspring produced by the daughters but not sons. These derivations show that the efficacy of mother-care scales with fecundity $m_i$, which goes to zero at late-life stages. In contrast, the efficacy of the long-term-cares and grandmother-care are cumulative in nature, they are therefore smaller at earlier life stages, get larger later, and do not go to zero even when one\textsc{\char13}s reproductive lifespan has ended. This allows an individual to be more contributive to its own reproductive success at late-life, thereby leading to strong selection in survival but not reproduction in late stages.

Interestingly, when mother-care is further introduced upon LTr+GM and LTs+GM, making them LTr+M+GM and LTs+M+GM, their extensive PRLS then disappears despite improved newborn survivorship (Fig. \ref{fig:xxxLifeTable}). This is because mothers outcompete grandmothers to caring for dependent infants, due to the higher chance for a mother to be with the care receiving infant than a grandmother. A care receiving infant thus relies on the investment in mid-life more than that in late-life, which makes investment in late-life become less relevant to one\textsc{\char13}s reproductive success and consequently terminates PRLS. 

How does the evolutionary path to PRLS look like? Our modelling framework can shed light on the properties of this trajectory. Here we approximate LTs+M+GM as the evolutionary starting point, assuming that agents invest in the offspring of their own and kins (represented by M and GM). This is supported by the observation of parental-care across numerous species \cite{Dulac2014}, and allomathering---caring for the offspring of neighbours---in primates \cite{fairbanks1990reciprocal}. We simulated the intermediate conditions between LTs+M+GM and LTs+GM. Our implementation of mother-care assumes that the presence of the mother reduces the ad-hoc death rate of a dependent infant by a factor $1/(1+\sigma)$. $\sigma$, the intensity of mother-care, is 9 by default, and we considered $\sigma=4, 1, 0.5, 0.2, 0.1$ in the intermediate conditions. Simulation on a condition was repeated five times, each lasted for 200,000 steps. We observed the gradual emergence of PRLS when reducing $\sigma$ from 9 to 0.2 (Fig. \ref{fig:xxxtransition}). There, PrR simultaneously increases from $0.11 \pm 0.02$ to $0.28 \pm 0.06$, and the stage of end-of-survival stage increases from $28.70 \pm 0.65$ to $34.27 \pm 0.94$. The stage of end-of-reproduction, $E$, is more intriguing: it increases from $23.93 \pm 0.94$ to $26.20 \pm 1.63$ at $\sigma=0.5$, and then declines to $24.30 \pm 1.84$. As we reduce $\sigma$ starting from 9, the survivorship of dependent infants goes down, and reproduction and somatic lifespan are extended to compensate for higher infant loss. But as $\sigma$ gets smaller, grandmother-care becomes more efficacious due to the extended somatic longevity. Thus, infant survivorship rebounces and reproductive lifespan shrinks. Somatic lifespan, however, does not shrink concurrently, because late-life investment has become efficarious and long somatic lifespan is beneficial. Hence, extensive PRLS emerges (Fig. \ref{fig:xxxtransition}). We emphasize that this trajectory is only a simplification of the reality. Withdrawal of mother-care leads to lower infant survival, which may not be favoured by selection and contradict our understanding that, the intermediate steps leading to a complex trait, such as extensive PRLS, need to be adaptive \cite{Pang2019Arose}. An adaptive transition may be achievable, for example, by introducing a trade-off of effort allocated to investing in descendants and other beneficial activities like foraging \cite{hawkes1997hadza}. In this way, a shift from investing in descendants to another activity could be beneficial to one\textsc{\char13}s reproductive success under certain circumstances.

To test the sensitivity of PRLS in our model, we limited the provision of reproduction-enhancing long-term-care to only the female offspring, and survival-enhancing long-term-care (LTs) to only the male offspring. The modified LTr+GM and LTs+GM are simulated five times, each lasted for 200,000 steps. Under these perturbations, PrR drops from $0.23 \pm 0.06$ to $0.20 \pm 0.09$ for LTr+GM, and from $0.28 \pm 0.09$ to $0.23 \pm 0.08$ for LTs+GM. The reduction of care-receiver weakens the PRLS signal, but the extensive PRLS nonetheless remains. Next, we simulated the condition GM and LTs+GM using a different beneficialness-to-deleteriousness ratio of mutation, $\eta=0.1, 0.9$ (default 0.5). $\eta=0.1$ (very deleterious mutation) leads to rapid extinction. At $\eta=0.9$ (mildly deleterious mutation), differential selection between survival and reproduction can be observed in both GM and LTs+GM. There, $r_i$ starts to diminish at $i \sim 30$, but $s_i$ stays close-to-one until $i \sim 50$ (Fig. \ref{fig:boxplot_sensitivity}). Their PrR, however, is $\ll 0.2$ despite the differential selection, because the weakly deleterious mutation allows $r_i$ to stay far above zero even without selection (Fig. \ref{fig:boxplot_sensitivity}). PrR may become >0.2 if we further introduce an additional locus to the genome that explicitly defines the reproductive lifespan and shuts down reproduction thereafter.

In summary, we have tested how different strategies of investment in descendants affect age-specific selection. We found that grandmother-care combined with long-term caring of descendants can give rise to extensive PRLS if mutation is considerably deleterious, or grandmother-care alone can suffice if mutation is very moderately deleterious. Our simulations also revealed a competitive relationship between mother-care and grandmother-care. In the earliest version of mother-care, it was posited that females turned off reproduction in mid-life \cite{williams1957pleiotropy}. Later, comparative study on humans and great apes showed that both have similar age of last-birth, which shifted our view from stop-reproduction-earlier to die-later \cite{Robbins2006,robson2006derived}. Our model simulation provided a more detailed description on how the investment strategies shifted the reproductive and somatic lifespan, and revealed a scaling law between fertility and infant survival. There are several other hypotheses to be explored, e.g., the tendency for males to choose younger female leads to the cessation of reproduction in mid-life \cite{Morton2013Mate}. We still have many questions left unanswered. Will somatic longevity significantly exceed reproductive longevity in other hypotheses after properly accounting for age-specific selection? What extra factors are necessary to make the evolutionary trajectory of PRLS adaptive? We leave these question to future studies.

We would like to thank Dario Riccardo Valenzano for valuable advice. This work is supported by SFB 1310 of German Research Foundation (DFG) awarded to TYP and Martin J Lercher, and Volkswagen Funding (VolkswagenStiftung) initiative``Life---A fresh scientific approach to the basic principles of life'' awarded to Martin J Lercher.

\nocite{*}

\bibliography{apssamp}

\end{document}